# Tuning of thermoelectric properties with changing Se content in $Sb_2Te_3$

D. Das[1], K. Malik[2], A. K. Deb[3], V.A. Kulbachinskii[4], V.G. Kytin[4], S. Chatterjee[5], D. Das[5], S. Dhara[6], S. Bandyopadhyay[1,7] and A. Banerjee[1,7,a)]

[1]Department of Physics, University of Calcutta, 92 A P C Road, Kolkata 700009, India

[2]Departmnt of Physics, Vidyasagar Evening College, 39 Sankar Ghosh Lane, Kolkata-700006, India

[3]Department of Physics, Raiganj University, Uttar Dinajpur 733134, India

[4]Department of Low Temperature Physics and Superconductivity, Physics Faculty, M.V. Lomonosov Moscow State University, 119991, Moscow, Russia

[5]UGC-DAE Consortium for Scientific Research, Kolkata Centre, Sector III, LB-8, Salt Lake, Kolkata 700098, India

[6]Surface and Nanoscience Division, Indira Gandhi Centre for Atomic Research, Kalpakkam 603102, India

[7]Center for Research in Nanoscience and Nanotechnology, University of Calcutta, JD-2, Sector-III, Salt Lake, Kolkata-700098, India

## ABSTRACT

Polycrystalline $Sb_2Te_{3-x}Se_x$ ($0.0 \leq x \leq 1.0$) samples were synthesized by the solid state reaction method. The structural analysis showed that up to the maximal concentration of Se, the samples possess the Rhombohedral crystal symmetry (space group $R\bar{3}m$). Increase of Se content increases the resistivity of the samples. Variation of phonon frequencies, observed from Raman spectroscopic study, depict anomalous behaviour around x = 0.2. The sample $Sb_2Te_{2.8}Se_{0.2}$ also shows maximum Seebeck coefficient, carrier concentration and thermoelectric power factor. Nature of scattering mechanism controlling the thermopower data has been explored. The thermoelectric properties of the synthesized materials have been analyzed theoretically in the frame of Boltzmann equation approach.

---

a) Author to whom correspondence should be addressed. Electronic mail: arbphy@caluniv.ac.in

## I. INTRODUCTION :

Thermoelectric (TE) effect refers to the phenomenon of direct conversion of heat to electric voltage and vice versa [1,2]. Efficiency of a TE material can be quantitatively expressed by the dimensionless term *figure of merit*, $ZT = \frac{S^2}{\rho \kappa} T$, where $S$, $\rho$ and $\kappa$, is respectively the Seebeck coefficient, electrical resistivity and thermal conductivity of the TE material and T is absolute temperature. By maximizing the Power-Factor (PF=$S^2/\rho$) and/or lowering the thermal conductivity ZT can be improved [3,4]. Antimony telluride ($Sb_2Te_3$) is a well known *p*-type TE material for near room temperature applications [5-8]. Incorporation of Se atoms into $Sb_2Te_3$ lattice modifies the nature of defect states, which in principle should lead to interesting changes in its TE properties. Efforts were thus devoted to study the effect of Se doping on structural [9], transport [10], electronic band structure [5] and TE [11] properties of $Sb_2Te_3$ alloys. Some anomalous behaviour were reported in the range of x = 0.0 – 1.0 for the $Sb_2Te_{3-x}Se_x$ system, which needs further attention. On the contrary, $Bi_2Te_3$ based chalcogenides, *viz.*, $Bi_2Te_{3-x}Se_x$ including the most compensated compound $Bi_2Te_2Se$ are well explored [12,13].

Here we investigate different compositions of $Sb_2Te_{3-x}Se_x$ ($0.0 \leq x \leq 0.1$) alloy. Room temperature powder X-ray diffraction (XRD) and thermal variation of resistivity, $\rho(T)$ data shows systematic variation with Se content. However, the *S*, *PF*, Hall carrier concentration of the charge carriers ($n_H$), and Raman spectroscopic study show some anomalous behaviour around x = 0.2. In this report, an attempt has been made to elucidate the origin of this anomalous behaviour. In addition, the *S(T)*, *ρ(T)*, $n_H(T)$ and *PF* data have also been theoretically simulated. The evaluation of band energy spectrum with Se content is predicted.

**II. EXPERIMENTAL:**

Polycrystalline $Sb_2Te_{3-x}Se_x$ (x = 0.0, 0.2, 0.6, 1.0) samples were synthesized by solid state reaction method [6]. The details of the structural characterization, $\rho(T)$, $S(T)$ measurements and room temperature Raman spectroscopic studies can be obtained elsewhere [6]. The temperature dependent Hall co-efficient $R_H(T)$ measurements were performed by Van-der-Pauw method on similar bar samples in a Closed Cycle Refrigerator (CCR) based 15 T magnet supplied by Cryogenic Ltd., UK.

**III. RESULTS AND DISCUSSION :**

Phase purity and structure of the $Sb_2Te_{3-x}Se_x$ ($0.0 \leq x \leq 1.0$) mixed crystals have been identified by XRD and the corresponding spectra are shown in fig. 1a. We have performed Rietveld refinement (utilizing *Materials Analysis Using Diffraction* [MAUD] program) using atomic positions and substitutions of all the synthesized $Sb_2Te_{3-x}Se_x$ (x=0.0, 0.2, 0.6, 1.0) samples. Space group $R\bar{3}m$ and point group $D_{3d}$ were used for the refinement [14-16]. The refinement parameters are provided in table 1. Figure 1b shows a typical refinement result for the $Sb_2Te_{2.8}Se_{0.2}$ sample.

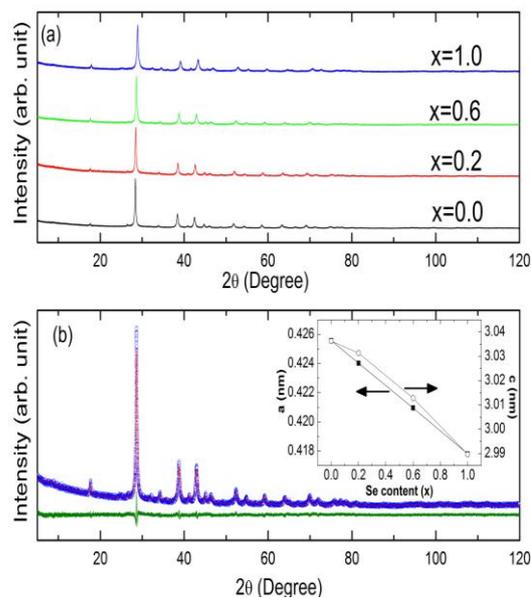

**Fig. 1:** (Color Online) (a) X-ray diffraction patterns of the samples $Sb_2Te_{3-x}Se_x$ (x=0.0, 0.2, 0.6, and 1.0). (b) X-ray diffraction pattern of sample $Sb_2Te_{2.8}Se_{0.2}$ after Rietveld refinement. The inset shows the variation of the lattice constants, '$a$' and '$c$', with Se composition for all the samples.

The variation of lattice constant with Se content for the samples are shown in inset. The linear contraction of lattice parameters with increased Se concentration closely follows the Vegard's law. The atomic radii of Te and Se are 142 pm and 100 pm, respectively [17]. Thus, substitution of Se at Te position should lead to decrease in the lattice parameter and hence the cell volume (Table-1). According to Vegard's law, the crystallographic parameters of a continuous substitutional solid solution vary linearly with concentration when the nature of the bonding is similar in the constituent phases. The XRD results thus confirm that synthesized $Sb_2Te_{3-x}Se_x$ alloys are single phase in nature and a complete solid solution has been formed with Se, substituting Te. However, close observation reveals that, the variation of lattice parameter '$a$' with Se concentration is exactly linear, i.e., strictly following Vegard's law. But, variation of lattice parameter '$c$' with Se content shows a little deviation from linearity around x=0.2. This might be related to the anomaly observed in $S(T)$, $R_H(T)$ or $n_H(T)$ and PF data of the $Sb_2Te_{2.8}Se_{0.2}$ sample discussed latter.

Figure 2 shows room temperature Raman spectra (RS) for all the synthesized samples. $Sb_2Te_3$ alloy exhibits four Raman active modes: $E_g^1$ (46 cm$^{-1}$), $A_{1g}^1$ (62 cm$^{-1}$), $E_g^2$ (113 cm$^{-1}$) and $A_{1g}^2$ (166 cm$^{-1}$) [18-20]. The RS (fig. 2) for Se-doped $Sb_2Te_3$ alloys depict three peaks at around 69 cm$^{-1}$, 112 cm$^{-1}$ and 166.6 cm$^{-1}$, that can be attributed to Raman active $A_{1g}^1$, $E_g^2$, and $A_{1g}^2$ vibrational modes, respectively. The active Raman mode $E_g^1$ (around 46 cm$^{-1}$) is out of the range measured in this work [6].

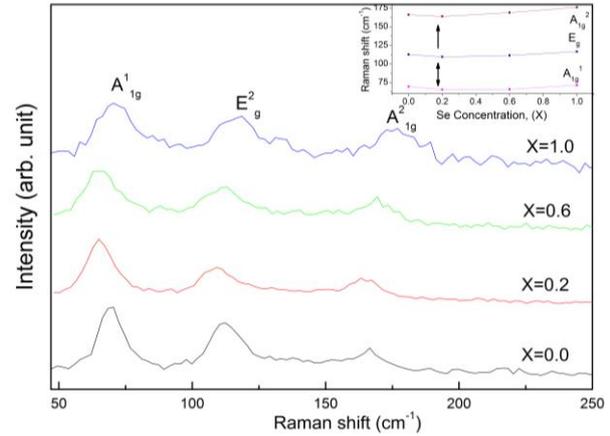

**Fig. 2:** (Color Online) Room temperature Raman spectra of the samples $Sb_2Te_{3-x}Se_x$ (x=0.0, 0.2, 0.6, and 1.0) recorded under excitation at λ=514.5 nm, indicating the presence of Raman active $A_{1g}^1, E_g^2, A_{1g}^2$ modes. Inset summarizes the observed Raman shifts of different vibrational modes with Se(x) content for the synthesized $Sb_2Te_{3-x}Se_x$ (0.0 ≤ x ≤ 1.0) samples.

Though Se atom is iso-electronic with Te, but is more electro-negative than Te. Incorporation of Se atoms in $Sb_2Te_3$ lattice will change its lattice dynamics. Inset in fig. 2 summarizes the observed Raman shifts of different vibrational modes with Se content and reveals that all the observed Raman active modes have very small shift with Se doping. Since the atomic masses of Sb and Te are comparable, the observed slight variation in $A_{1g}^1$, $E_g^2$ and $A_{1g}^2$ modes with Se (x) content is anticipated for $Sb_2Te_{3-x}Se_x$ (0.0 ≤ x ≤ 1.0) mixed crystals, where Se concentration is not significantly higher. However, close observation of fig. 2 (inset) divulges that, initially for low Se content (x=0.2) the phonon frequencies shift to lower wave number side. But with further increase of Se i.e., for x = 0.6 and 1.0 observed phonon frequencies gradually shift to higher wave number side. The unit cell of $Sb_2Te_3$ - like compounds has five quintuple layers ($Te^1$-Sb-$Te^2$-Sb-$Te^1$) weakly bound by Van-der-Waals force in which Te atoms exhibit two different environments *i.e.*, $Te^1$ and $Te^2$. $Te^2$ atoms are the centre of mass of lattice vibration [12] and Raman active modes directly manifest the vibrational properties of Sb-Te(2)/Te(1) bonds. Sb-$Te^2$ bond is more polar than Sb-$Te^1$ bond. Initially for low concentration (x=0.2), Se preferentially replaces Te at $Te^2$ site, and subsequently for higher concentration (x=0.6, 1.0) the Se atoms continue to replace Te of $Te^1/Te^2$ sites in a random manner [21]. This might lead to the observed anomaly in phonon frequency for the $Sb_2Te_{2.8}Se_{0.2}$ sample (inset in fig. 2, marked with an arrow).

The ρ(T) data of the polycrystalline $Sb_2Te_{3-x}Se_x$ samples depicts that ρ increases with increasing Se concentration (Figure 3a). Pristine $Sb_2Te_3$ always possesses over-stoichiometric Sb atoms along with native point defects, viz., $V_{Te}$ and segregated Te [6,22,23]. The over-stoichiometric Sb atoms occupy prevailingly the $Te^2$ sites in the Te sublattice, giving rise to AS defects of $Sb_{Te}$ type. Due to such AS defects, $Sb_2Te_3$ always shows a *p*-type conductivity with hole concentration around $10^{20}$ cm$^{-3}$[9]. With increasing Se concentration the formation energy of these AS defects increases. As a consequence the formation probability of AS defects decreases, which in turn leads to the decrease of carrier (hole) concentration in $Sb_2Te_{3-x}Se_x$ samples with increasing x. Thus ρ increases with increasing Se content for the reported $Sb_2Te_{3-x}Se_x$ alloys.

Figure 3a further depicts that, while $Sb_2Te_3$ and $Sb_2Te_{2.8}Se_{0.2}$ samples exhibit weakly metallic ρ, but an activated ρ(T) behaviour is observed for higher Se content samples. The observed metallic ρ(T) data, arising due to increase of intrinsic carrier concentration at high temperatures, is typical for these heavily doped narrow band semiconductors [24]. In order to extract the nature of carrier scattering in $Sb_2Te_{3-x}Se_x$ (x=0.0, 0.2) samples, $\rho(T)$ curve is fitted with the power-law expression $\rho=\rho_o+AT^n$. For both the samples, obtained best fit value of $n$ is 1.66. Recently Dutta *et al.* reported same $n$ value for $Sb_2Te_3$ crystals [8]. On the other hand, similar value of $n$ (=1.3) for $Sb_2Te_{3-x}Se_x$ system was also reported earlier by Kulbachiniskii *et al.* [5]. For bulk $Sb_2Te_3$ and related TI systems, transport properties of surface state are often mixed with bulk state, which probably gives rise to the unusual value of the exponent in the low Se content $Sb_2Te_{3-x}Se_x$ samples [6,8].

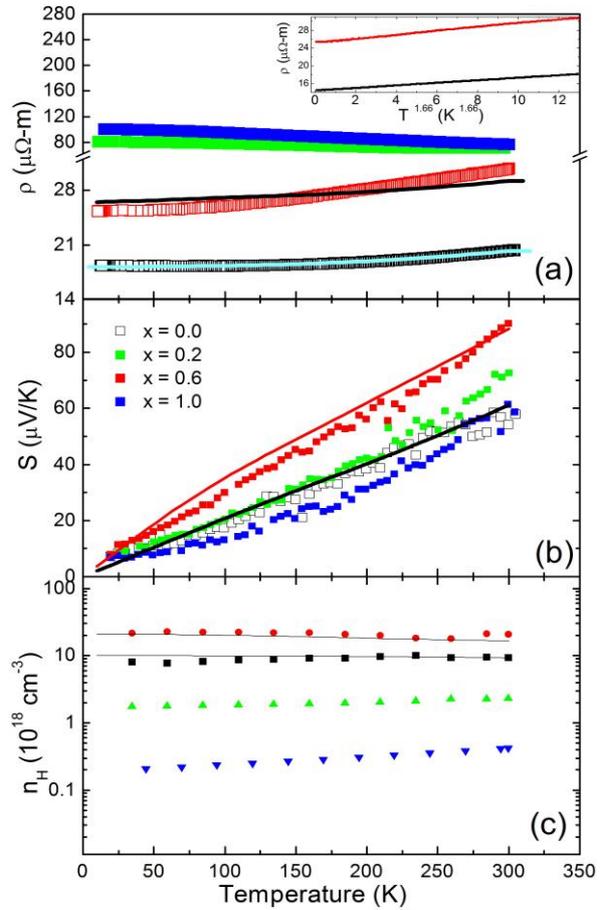

**Fig. 3:** (Color online) Experimental (points) and simulated (solid lines) temperature dependencies of (a) the electrical resistivity $\rho$; (b) the Seebeck coefficient $S$; and (c) the Hall carrier concentration $n_H$ for the $Sb_2Te_{3-x}Se_x$ (x=0.0, 0.2, 0.6, and 1.0) samples. The simulation was performed simultaneously (with the same parameters) in the frame of Boltzmann equation approach for the samples with x = 0.0 and 0.2 Se content. Figure 3a(inset) demonstrates the power-law fit, $\rho = \rho_0 + AT^n$ with n~1.66 to the ρ(T) data for $Sb_2Te_3$ and $Sb_2Te_{2.8}Se_{0.2}$ sample.

The *S(T)* data depicts that all the reported samples are *p*-type in nature (Figure 3b). *S(T)* initially increases with increasing x from x=0.0 to 0.2. But for x ≥ 0.2, *S(T)* decreases. Compositional dependence of scattering parameter (r) is estimated from the *S(T)* data. For a heavily doped semiconductor (for a single valley in isotropic case), *S* is given by [13]:

$$S = \pm \frac{k_B}{e} \left[ \eta_F - \frac{(r+5/2)F_{r+3/2}(\eta_F)}{(r+3/2)F_{r+1/2}(\eta_F)} \right] \quad (1)$$

Here $\eta_F = E_F/(k_B T)$ is reduced Fermi energy; parameter $r$ describes the energy dependence of scattering time, and

$$F_n = \int_0^\infty d\eta \frac{\eta^n}{1+\exp(n-n_F)} \quad (2)$$

is Fermi integral. Now, $r = -0.5$ corresponds to acoustic phonon scattering, $r = 0.5$ is scattering due to optical phonons and $r = 1.5$ denotes scattering by ionized impurities. We estimate the Fermi energy ($E_F$) in two samples, *viz.*, $Sb_2Te_3$ and $Sb_2Te_{2.8}Se_{0.2}$, exhibiting metallic behaviour by a simple model

$$E_F = \frac{\hbar^2}{2m^*} \left[ 3\pi^2 n_H \right]^{2/3}, \quad (3)$$

where $n_H$ is carrier concentration (fig. 3c). Reasonable values of $E_F \approx 100$ *meV* and *120 meV*, are obtained respectively for x=0.0 and 0.2 samples [25,26]. Using these Fermi-energies and formulae 1 and 2 we may estimate the scattering parameter $r$. Such procedure yields $r = -0.5$ and $r = 0.1$, respectively for x = 0.0 and x=0.2 alloys. The obtained best fit value of $r = -0.5$ for $Sb_2Te_3$ alloy (i.e., x = 0.0) corroborates with the reported results [13]. The *S(T)* data thus reveals that in SbTeSe based degenerated semiconductors, increasing of Se content shift the scattering from preferentially acoustic phonon closer to impurity scattering.

For $Sb_2Te_{2.4}Se_{0.6}$ and $Sb_2Te_2Se$ samples, the activation energy ($E_{act}$) is estimated from the $\rho(T)$ data using the relation:

$$\rho = \rho_0 \exp\left(\frac{E_{act}}{2k_B T}\right) \quad (4)$$

where, $\rho_0$ is a constant. The estimated $E_{act}$ values, 9.24 meV and 12.30 meV respectively for $Sb_2Te_{2.4}Se_{0.6}$ and $Sb_2Te_2Se$, indicate that estimated transport gap ($E_{act}$) is much smaller than the reported band gap ($E_g$) [27]. $E_g$ arises due to the difference between the lower conduction band (LCB) and upper valence band (UVB) [22]. However, according to Arkap *et al.*, while $E_g$ is determined by band structure, $E_{act}$ is actually linked to the presence of point defects [13]. $Sb_2Te_3$ hosts various kinds of defects, *viz.*, AS defects of $Sb_{Te}$ type, $V_{Te}$ and segregated Te. Increasing Se content in $Sb_2Te_3$ matrix enhances the interplay between these defects with Se atoms initially replace $Te^2$ atoms, but for higher concentration it continue to replace $Te^1/Te^2$ atoms randomly. This might lead to observed change from metallic to activated behaviour in $\rho(T)$ data for x > 0.2. The energy state due to the native defects in $Sb_2Te_{3-x}Se_x$ system probably lays between LCB-UVB and leads to the smaller $E_{act}$ in the activated samples. Further, when the composition is tuned from $Sb_2Te_3$ to $Sb_2Te_2Se$, concentration of hole in the system decreases. Around the composition x=1 ($Sb_2Te_2Se$), Se/Te sublattice is expected to be ordered with almost complete compensation of donor and acceptor impurities [13]. Therefore around x=1 bulk conductivity should be minimum, which is correctly reflected in our $\rho(T)$ data [see fig. 3a].

Figure 3c represents the thermal variation of $n_H$ data for all the $Sb_2Te_{3-x}Se_x$ samples, measured in a magnetic field of 12 T. Like $S(T)$ data initially with increasing x, $n_H$ increases for $Sb_2Te_3$ and $Sb_2Te_{2.8}Se_{0.2}$, but for x ≥ 0.2, $n_H$ decreases, indicating that apparent hole concentration is highest in $Sb_2Te_{2.8}Se_{0.2}$. Here we would like to represent a plausible explanation for such anomalous behaviour of $n_H$ in $Sb_2Te_{3-x}Se_x$. Band structure calculation indicates that, the UVB of $Sb_2Te_3$ consists of six ellipsoids and the lower valence band (LVB) is known to be multivalleyed [5,24]. While explaining the Shubnikov-de Haas (SdH) and transient thermoelectric effect (TTE) data for $Sb_2Te_{3-x}Se_x$ (0.0 ≤ x ≤ 1.0) system Kulbachinskii *et al* indicates the presence of another valence band (NVB) for x ≥ 0.2 and clearly predict two different regions *viz.*, 0 ≤ x ≤ 0.2 and 0.2 < x ≤ 1.0 in band structure [5]. In the first region 0 ≤ x ≤ 0.2, only UVB and LVB contribute. But in the second region 0.2 <

$x \leq 1.0$ the contribution comes from UVB, LVB and NVB, where NVB moves up and UVB moves down due to Se doping. It should be mentioned that, UVB and LVB is not equally populated, the ratio of the density of holes in LVB to that of UVB is around 390 [28]. In view of this large ratio, LVB contributes mostly in the conduction mechanism. The evaluation of the band structure with Se content in $Sb_2Te_{3-x}Se_x$ ($0.0 \leq x \leq 1.0$) alloy as predicted by Kulbachinskii *et al.* [5], and correspondingly the net contribution of carrier from UVB, LVB and NVB might be related to the experimentally observed $n_H$ behaviour.

In the framework of Boltzmann equation approach, we simulated simultaneously (with the same parameters) $\rho(T)$ (see fig. 3a), $S(T)$ (see fig. 3b), and $n_H(T)$ (see fig. 3c) for the samples with x=0.0 and 0.2. Theoretical dependencies are shown in fig. 3a-c by solid lines. However we were unable to reproduce small value of Seebeck coefficient in samples with x=0.6 and x=1.0 Se content without unrealistic change of band parameters. We suggest that part of holes in material is localized and to fit data for samples with high Se content the localization of part of holes should be taken into account. The theoretical model and method of calculation is described in Ref. 5. In the fitting following scattering mechanisms were taken into account: Acoustic phonon scattering, ionized acceptors scattering, grain boundary scattering. All acceptors were assumed to be ionized. Other parameters, *viz.*, effective masses, band edges, deformation potential, acceptor concentration etc. are taken from the literature. For the fitting of sample with x=0.2 Se content only position of second valence band, grain size and acceptor concentration were changed with respect to $Sb_2Te_3$. The details of the simulated parameters are provided in table 2. The optimal fitting value for the distance between valence bands is larger than was obtained experimentally earlier [28]. This is probably due to localization of the part of holes in investigated samples. This could be also origin of the very small Seebeck coefficient in the samples with x=0.6 and x=1.0. For these samples (x = 0.6, 1.0) localization of the part of holes can explain the $\rho(T)$ data, which has an activated character (see fig. 3a). These samples have too large value of the Hall coefficient and small value of Seebeck coefficient. One of the reasons may be (probably) due to modification of band structure with high Se content [5] and localization of the part of holes.

Figure 4 shows the compositional dependence of PF for $Sb_2Te_{3-x}Se_x$ samples estimated from the measured quantities. Theoretical values of PF for $Sb_2Te_3$ and $Sb_2Te_{2.8}Se_{0.2}$ alloys, calculated from simulated $S(T)$ and $\rho(T)$ data, is also plotted in fig. 4. The PF increases with increasing temperature and reveals non saturating behaviour around room temperature. Similar to the behaviour observed for $S(T)$ and $n_H(T)$ data, maximum value of PF is also observed for the $Sb_2Te_{2.8}Se_{0.2}$ alloy. This further indicates that, increasing Se content in $Sb_2Te_3$ do not always contribute a positive role in increasing the thermoelectric performances of $Sb_2Te_{3-x}Se_x$ samples. However conclusive evidence can be drawn only after estimating ZT. It should be recalled that, $Sb_2Te_{2.8}Se_{0.2}$ possess highest $n_H$ (~$10^{19}$/cc). It is thus quite justified to assume that tuning of carrier concentration as well as band structure engineering including the optimization of band parameters in $Sb_2Te_{3-x}Se_x$ system should play a crucial role in obtaining a good TE material.

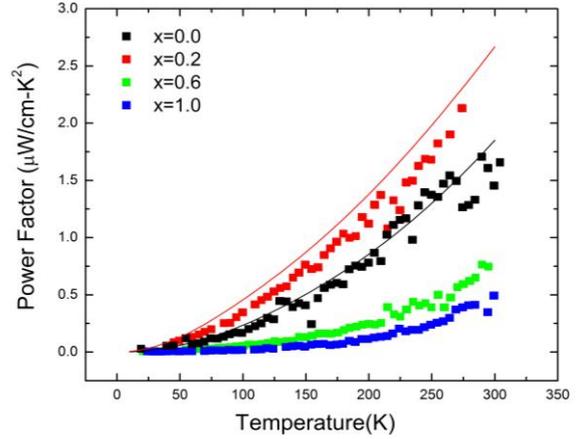

**Fig. 4:** (Color online) Thermal variation of thermoelectric Power Factor for the $Sb_2Te_{3-x}Se_x$ (x=0.0, 0.2, 0.6, and 1.0) samples. Points denote experimental data and solid lines represent theoretical fit.

### IV. CONCLUSION :

XRD data and Raman spectra confirm that all the synthesized $Sb_2Te_{3-x}Se_x$ samples exhibit rhombohedral crystal geometry. It has been realized that the surface states are often mixed with the bulk state, giving rise to the observed metallicity in $Sb_2Te_3$ and $Sb_2Te_{2.8}Se_{0.2}$ samples. Transport gap ($E_{act}$), estimated in high Se content samples depicting activated behavior, is smaller than reported band gap ($E_g$) probably due to the presence of point defects and the tail of the density of state. Acoustic phonon scattering dominates the *S(T)* data for the synthesized host $Sb_2Te_3$. However, with increasing Se concentration impurity scattering gradually dominates the *S(T)* data. Theoretical calculation, based on four-band model, demonstrates that position of the second valance band as well as acceptor concentration are sensitive to Se concentration and part of holes in high Se content samples are localized.

**ACKNOWLEDGEMENT :**

The work is supported by DST, Govt. of India-RFBR, Govt. of Russia (DST Ref. no.: INT/RUS/RFBR/P-183, RFBR grant: IND-a 15-52-45037). The financial grant, including the fellowship of first author DD, received from UGC-DAE Consortium for Scientific Research, Kalpakkam node (Project reference no.:CSR-KN/CRS-65/2014-15/505) is also gratefully acknowledged. The authors would also like to thank DST, Govt. of India, for low temperature high-magnetic-field facilities at UGC-DAE CSR, Kolkata Centre.

**Table 1:** Rietveld refinement parameters, *viz.*, lattice parameters, unit cell volume, position coordinates, site occupancy, Debye-Waller factor ($B_{iso}$), reliability parameters [$R_w$(%), $R_b$(%), and $R_{exp}$ (%)], and goodness of fit (Gof or $\chi^2$) value, as obtained using MAUD software for the samples $Sb_2Te_{3-x}Se_x$ (x=0.0, 0.2, 0.6, and 1.0). The corresponding values of the estimated errors are also provided.

| Phase | $Sb_2Te_3$ [$R\bar{3}m$] | $Sb_2Te_{2.8}Se_{0.2}$ [$R\bar{3}m$] | $Sb_2Te_{2.4}Se_{0.6}$ [$R\bar{3}m$] | $Sb_2Te_2Se$ [$R\bar{3}m$] |
|---|---|---|---|---|
| Cell (Å) | a: 4.2558 (4.3 x $10^{-5}$)<br>c: 30.3629 (7.4 x $10^{-4}$) | *a*: 4.2403 (1.2 x $10^{-4}$);<br>*c*: 30.3134 (1.1 x $10^{-3}$) | *a*: 4.2096 (1.1 x $10^{-4}$);<br>*c*: 30.1278 (1.1 x $10^{-3}$) | *a*: 4.1781 (1.6 x $10^{-4}$);<br>*c*: 29.8964 (1.6 x $10^{-3}$) |
| Cell volume | 476.25 | 472.02 | 462.36 | 451.97 |
| $Sb_x$/$Sb_y$/$Sb_z$ | 0.0 / 0.0/ 0.3994 (4.1x$10^{-5}$) | 0.0 / 0.0/ 0.6027 (6.2x$10^{-5}$) | 0.0 / 0.0/ 0.3962 (4.3x$10^{-5}$) | 0.0 / 0.0/ 0.3943 (4.7x$10^{-5}$) |
| $Te1_x$/$Te1_y$/$Te1_z$ | 0.0 / 0.0/ 0.7874(2.4x$10^{-5}$) | 0.0 / 0.0/ 0.7879(4.0x$10^{-5}$) | 0.0 / 0.0/ 0.7867(2.9x$10^{-5}$) | 0.0 / 0.0/ 0.7860(3.5x$10^{-5}$) |
| $Se1_x$/ $Se1_y$/ $Se1_z$ | ------- | Te2/ Se1: 0.0/ 0.0/ 0.0 | Te2/ Se1: 0.0/ 0.0/ 0.0 | 0.0/ 0.0/ 0.0 |
| $B_{isoSb/Te1/Se1}$ | Sb: 2.619 (0.026)<br>Te1: 1.522 (0.032)<br>Te2: 0.147 (0.029) | Sb: 1.1917 (0.021)<br>Te1: 1. 1917 (equal),<br>Te2/Se1: 1.1917 (equal) | Sb: 1.8722 (0.020)<br>Te1: 1. 8722 (equal),<br>Te2/Se1: 1. 8722 (equal) | Sb: 1.544 (0.024)<br>Te1: 1.544 (equal),<br>Se1: 1.544 (equal) |
| $R_w$ (%) | 3.941 | 5.967 | 5.171 | 4.814 |
| $R_b$(%) | 3.091 | 4.617 | 4.089 | 3.819 |
| $R_{exp}$(%) | 2.104 | 4.602 | 4.115 | 3.886 |
| Gof or $\chi^2$ | 1.873 | 1.296 | 1.257 | 1.239 |

**Table 2:** The simulated parameters of the investigated $Sb_2Te_3$ and $Sb_2Te_{2.8}Se_{0.2}$ alloys obtained by fitting the experimental temperature dependencies of the Seebeck coefficient, the resistivity, and the Hall carrier concentration.

|  |  |  | $Sb_2Te_3$ [ $R\bar{3}m$ ] | $Sb_2Te_{2.8}Se_{0.2}$ [ $R\bar{3}m$ ] |
|---|---|---|---|---|
| Bandgap (eV) |  |  | 0.25 | 0.25 |
| Second Valance Band (eV) |  |  | 0.19 | 0.08 |
| Effective Masses ($m_0$) | Light Holes |  | 0.043 | 0.043 |
|  | Heavy Holes |  | 0.15 | 0.15 |
| Band Extrema positions (b) | Light Holes | b1 | 0.705 | 0.705 |
|  |  | b2 | 0.615 | 0.615 |
|  |  | b3 | 0.615 | 0.615 |
|  | Heavy Holes | b1 | 0.534 | 0.534 |
|  |  | b2 | 0.341 | 0.341 |
|  |  | b3 | 0.341 | 0.341 |
| Acceptor Concentration ($cm^{-3}$) |  |  | $2.0 \times 10^{19}$ | $2.3 \times 10^{19}$ |
| Scattering Parameters | Density ($Kg/m^3$) |  | 6500 | 6500 |
|  | Sound Velocity (m/s) |  | 1780 | 1780 |
|  | Deformation Potential (eV) |  | 3 | 3 |
| Grain Boundaries | Grain Size (nm) |  | 15.3 | 10 |
|  | Scattering Constant |  | 0.3 | 0.3 |
| Dielectric Constant |  |  | 101 | 101 |